\documentclass[journal]{IEEEtran}

\DeclareTextSymbol{\degre}{T1}{6}
\DeclareTextSymbol{\degre}{OT1}{23}
\usepackage{etex}
\usepackage{url}
\usepackage[cmex10]{amsmath}
\usepackage{amsfonts}

\usepackage{dsfont}
\usepackage{epstopdf}
\usepackage{graphicx}
\usepackage[noadjust]{cite}
\usepackage{pstricks}
\usepackage{pstricks-add}
\usepackage{pst-plot}
\usepackage{tikz}
\usepackage{pgfplots}
\usetikzlibrary{plotmarks}
\usetikzlibrary{shapes,arrows}
\usepackage{bbold}
\usepackage{subfigure}

\begin{document}
\title{Introducing Decentralized EV Charging Coordination for the Voltage Regulation}

\author{Olivier~Beaude,~\IEEEmembership{Student Member,~IEEE,}
				Yujun~He,~\IEEEmembership{Student Member,~IEEE,}
        and ~Martin~Hennebel~
\thanks{O. Beaude PhD student at L2S, Supelec Power System Department and Renault SAS, in Paris, France, Y. He PhD student at Supelec Power System Department, M. Hennebel researcher and professor at Supelec Power System Department, E3S, in Gif-sur-Yvette, France. E-mail: olivier.beaude@renault.com.}
\thanks{Manuscript received June 19, 2013
}}

\markboth{ISGT Copenhagen 2013, October 2013}%
{Shell \MakeLowercase{\textit{et al.}}: Bare Demo of IEEEtran.cls for Journals}

\maketitle

\begin{abstract}
This paper investigates a decentralized optimization methodology to coordinate Electric Vehicles (EV) charging in order to contribute to the voltage control on a residential electrical distribution feeder. This aims to maintain the voltage level in function of the EV's power injection using the sensitivity matrix approach. The decentralized optimization is tested with two different methods, respectively global and local, when EV take into account their impact on all the nodes of the network or only on a local neighborhood of their connection point. EV can also update their decisions asynchronously or synchronously. While only the global approach with asynchronous update is theoretically proven to converge, using results from game theory, simulations show the potential of other algorithms for which fewer iterations or fewer informations are necessary. Finally, using Monte Carlo simulations over a wide range of EV localization configurations, the first analysis have also shown a promising performance in comparison with uncoordinated charging or with a "voltage droop charging control" recently proposed in the literature.
\end{abstract}


\IEEEpeerreviewmaketitle

\section{Introduction}

\IEEEPARstart{V}{oltage} regulation is one of the significant ancillary services  \cite{Geth2012}  in distribution systems. In the evolution towards a "smarter grid", it has to become more flexible to deal with the variation of consumer's need and distributed generations \cite{Baran2007}.   

In this context, smart grid is envisioned to make the most of potential interactions between power systems and electric vehicles (EV). A large part of literature has been devoted to a centralized approach (see \cite{Clement-Nyns2009}) to perfectly schedule EV charging according to various objectives (power losses, voltage deviations, charging costs...) while the behavior of end users is less considered. Thus, a decentralized approach could contribute to the further development of practical coordination mechanisms, the next step before real implementation. \cite{Leemput2011} recently gave an overview of smart mechanisms explored in EV smart charging literature comparing centralized and decentralized results. Some of the distributed methods leading to promising results are based on game theory, which is a powerful tool to study their properties \cite{Saad2012}. This comes from the fact that Nash equilibria may be attractors for many distributed mechanisms designed in coordination problems. Consequently, these equilibria, and particularly the study of their efficiency, may play a major role in this context. Recent literature in this field contains \cite{Gan2012} which optimizes the interaction between a transformer and a group of EVs and \cite{Wu2012} in the context of wind power integration. To the best of our knowledge, our work is the first to apply this framework to the issue of voltage control.

\section{Problem description}

\subsection{EV charging modeling}
 
The availability of EV charging is concerned with a wild research on user's behavior. In this work, it will be considered that most of users park their car at home during the night hours and require that it is enough charged to travel next day.  The EV charging in this context can be simply modeled as a controllable power-constant load in the band of $[0, P_{max}]$, where $P_{max}$ is the maximal charging power of EV. The value of $P_{max}$ is varied from $3$kW to $48$kW, depending on different technologies of EV charging. With the state-of-charge (SoC) representing the charging state of EV batteries, the constraints of charging can be written as
\begin{equation}
 SoC_{\textrm{min}} \leq SoC_{\textrm{init}}+\sum_{t=0}^{T}P_{t} \leq SoC_{\textrm{max}} \text{ ,}
\end{equation}
where $SoC_{\textrm{init}}$ is the initial value when EV parks at home, $SoC_{\textrm{min}}$ is the minimal acceptable value for the user's next day traveling, $SoC_{\textrm{max}}$ the maximal value limited by the battery of EV and $P_{t}$ the EV charging power during time slot $t$.

Given the SoC at time $t$, and $SoC_{\textrm{min}}$ (respectively $SoC_{\textrm{max}}$), the minimal (respectively maximal) charging power at time $t$, denoted by $\underline{P}_{t}$ (respectively $\overline{P}_{t}$), can be calculated, providing  
\begin{equation}
 \underline{P}_{t} \leq  P_{t} \leq  \overline{P}_{t} \text{ .}
\end{equation}

As known, cables on distribution systems have a great R/X ratio (close to 1).  Hence the active power delivery to EV chargers can generate voltage drops, and a charging power modulation can contribute to the voltage control, as is highlighted hereafter. In the following, time indexes will be omitted given that the proposed methodology is repeated at each time slot.

\subsection{Voltage control with a sensitivity analysis}

The sensitivity analysis \cite{VanCutsem1998} is used to evaluate the changes of some quantity $\eta$ of interest if changes of some parameter $p$ take place in electric systems. In this work, the changes of bus voltage magnitude $\Delta V$ will be evaluated and the parameters of concern are bus power injections $P,Q$.

The concerned sensitivity matrix comes from the network's load-flow equations as follows:
\begin{equation}
\label{LFMatrix}
\begin{cases}
P_{i}= \displaystyle \sum_{j=1}^{N} V_{i}V_{j}Y_{ij}\cos(\delta_{i}-\delta_{j}-\theta_{ij}) \\
Q_{i}= \displaystyle \sum_{j=1}^{N} V_{i}V_{j}Y_{ij}\sin(\delta_{i}-\delta_{j}-\theta_{ij}) 
\end{cases}
\end{equation}
where $P_{i}$, $Q_{i}$, $V_{i}$ and $\delta_{i}$ are respectively the active and reactive power injection, bus voltage magnitude and angle at bus $i$; $Y_{ij}$ and $\theta_{ij}$ are respectively the module and argument of the element $(i,j)$ of the network admittance matrix.  

By calculating the partial derivatives of (\ref{LFMatrix}), its Jacobian matrix is obtained:
\begin{equation}
\begin{bmatrix}
\Delta P \\
\Delta Q
\end{bmatrix}
=
\displaystyle
\begin{bmatrix}
\frac{\partial P}{\partial \delta} \frac{\partial P}{\partial V} \\
\frac{\partial Q}{\partial \delta} \frac{\partial Q}{\partial V} \\
\end{bmatrix}
\begin{bmatrix}
\Delta \delta \\
\Delta V \\
\end{bmatrix}
\end{equation}
The coupling of $V$-$P$ and $V$-$Q$ can be expressed from the Jacobian matrix. In \cite{He2012}, we have 
\begin{equation}
\Delta V_{p}=S_{V_{p},P_{c}}\Delta P_{c}+S_{V_{p},Q_{c}}\Delta Q_{c}
\end{equation}
where $p$ is the notation of the set of pilot nodes whose voltage profile should be maintained, while $c$ is the notation of the set of nodes where load injection is controlled. The matrices $S_{Vp,Pc}$ and $S_{Vp,Qc}$ are called sensitivity matrices respectively for the coupling $V$-$P$ and $V$-$Q$.

Considering active power control is concerned in EV charging, only the matrix $S_{Vp,Pc}$ will be used for this study. The charger converter could also allow a reactive power control. 

Before presenting the decentralized algorithms to control the voltage, a key function is introduced

\begin{equation}
f_V(V_p)=
\begin{cases}
(V_p-\underline{V})^2 \text{ if } V_p<\underline{V} \\
0 \text{ if } \overline{V} \leq V_p \leq \underline{V} \\
(V_p-\overline{V})^2 \text{ if } V_p>\overline{V} \\
\end{cases}
\end{equation}

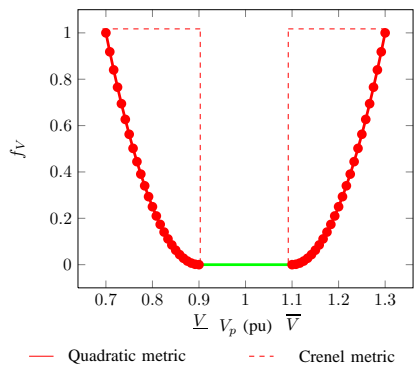
\begin{figure}[!htbp]
\centering
\scalebox{0.65}{
\begin{tikzpicture}
\begin{axis}[xlabel=$V_p$ (pu),ylabel=$f_V$]

\addplot[domain=0.9:1.1,green,ultra thick]{0};
\addplot[domain=1.1:1.3,mark=*,red,ultra thick]{(x-1.1)^2/0.04};
\addplot[domain=0.7:0.9,mark=*,red,ultra thick]{(x-0.9)^2/0.04};

\end{axis}
\draw[red,dashed] (0.5,5.3)--(2.5,5.3);
\draw[red,dashed] (2.5,0.5)--(2.5,5.3);
\draw[red,dashed] (4.3,0.5)--(4.3,5.3);
\draw[red,dashed] (4.3,5.3)--(6.3,5.3);
\draw[red,dashed] (3.5,-1.4)--(4,-1.4);
\node at (5.5,-1.4) {Crenel metric};
\draw[red] (-1,-1.4)--(-0.5,-1.4);
\node at (1,-1.4) {Quadratic metric};
\node at (2.5,-0.7) {$\underline{V}$};
\node at (4.4,-0.7) {$\overline{V}$};
\end{tikzpicture}
}
\caption{Objective concerning voltage regulation}
\label{fig:PhysMetr}
\end{figure}

This will permit to quantify how efficient is the voltage regulation. In practice, this must be determined according to the penalties paid by the Distribution Network Operator (DNO) to keep the voltage within its standard limits. With this definition, while the voltage is between its standard limits ($0.9$ and $1.1$ pu), the DNO has no penalty, and these penalties are quadratic out of this interval. As a comparison and as presented in dashed in Fig.\ref{fig:PhysMetr}, a second metric will be considered here, called "crenel" function: $0$ between $0.9$ and $1.1$ pu, $1$ otherwise. This is a first step to analyze the sensibility of the results to the metric used.

\subsection{Decentralized algorithm for voltage control}

The decentralized algorithm used in this work is an iterative algorithm which is called the \textit{best response dynamics} (BRD) in game theory \cite{Fudenberg1991}. This implements a \textit{communication phase taking place off-line}, before charging begins, to coordinate charging decisions of all the EV connected to the same network. Note that an online application of the proposed methodology could also be considered : if the charging at time $t$ has already begun but if there is a need for updating the charging decisions (for example, a new EV has just connected to the network) before time $t+1$, the decentralized process could be applied again, having updated the charging needs of all the EV which were charging at this time. As soon as a new charging configuration is obtained, then it is applied.

By default, without knowing the state of the voltage on the pilot nodes, each EV (with an automaton) initially chooses a charging power (for example $ P_{rated}$). Receiving all the EV charging decisions, an \textit{aggregator} calculates the voltage on all the pilot nodes and feedbacks EV with this information. Therefore each EV updates its charging decision to minimize an objective and reports this change to the aggregator. This procedure is then repeated while a stopping criterion is not reached.

Observe that EV can update their decisions \textit{synchronously} or \textit{asynchronously} (EV $1$ updates its choice, then the aggregator calculates and send the pilot nodes' voltage to all the EV, then EV $2$ updates and reports its charging power...). Once this communication phase is finished, each EV knows its charging power. 

Using the sensitivity matrix, two decentralized approaches for the voltage regulation will be distinguished according to the objective used by EV to update their charging decisions. In the first one, all EV follow the same objective which concerns all pilot nodes. Setting $\Delta P_{i}$ and supposing $\Delta P_{-i}=(\Delta P_{1},\Delta P_{2},...,\Delta P_{i-1},\Delta P_{i+1}...,\Delta P_{I})$ fixed, EV $i$ minimizes
\begin{equation}
f^{global}(\Delta P_{i},\Delta P_{-i})=\displaystyle \sum_{p=1}^{N_{p}} f_V(V_{p}-V_{\textrm{ref}}+\displaystyle \sum_{c}S_{V_{p},P_c} *\Delta P_c)
\end{equation}
where $V_{p}$ is the actual voltage measurement and $V_{\textrm{ref}}$ the setpoint for voltage control. 

 
In the second one, EV $i$ is more particularly concerned with the voltage on its neighborhood, denoted by $\mathcal{V}_{i}$, and defined by the electrical network topology (typically, a single feeder or a part of this feeder) given that its charging choice can more directly influence the state of these nodes. To update its charging choice, EV $i$ will then minimize
\begin{equation}
f^{local}_i(\Delta P_{i},\Delta P_{-i})=\displaystyle \sum_{p \in \mathcal{V}_{i}} f_V(V_{p}-V_{\textrm{ref}}+\displaystyle \sum_{c}S_{V_{p},P} *\Delta P)
\end{equation}

It should be noted that one key advantage of the local approach in comparison to the global one is that only the local voltage state is needed for EV $i$ to take its decision while in the global case, the voltage on all the pilot nodes must be sent, which makes a bigger need for data exchange.

\subsection{Convergence of the global asynchronous approach inherited from the game theoretical class of potential games}

This part describes the link between the decentralized charging algorithms defined previously and a field of game theory called potential games (see \cite{beaude12} for more details). The motivation for introducing here the tools from non-cooperative game theory is as follows. The practical scenario considered here corresponds to the situation where each EV owner decides when to charge his vehicle. In such a scenario, the variable $P_i$ is controlled by EV $i$ only. This is therefore a distributed optimization problem. One of the powerful links between distributed optimization and game theory is that scenarios involving several individual optimizers which update their configuration over time may converge to an equilibrium of a certain game. This is the reason why we define here an \textit{auxiliary game} of interest and then describe the properties of the distributed algorithm of this work, coming from the particular structure of this game. This auxiliary game under strategic form consists of three main components~:

\begin{itemize}
\item \textit{Players}: EV $i \in \mathcal{I}=\left\{1;I\right\}$ connected to the district ;
\item \textit{Actions}: charging power $P_{i}$, or equivalently $\Delta P_i$. $P_{-i}$ are the actions of $\mathcal{I} \backslash \left\{i\right\}$, denoting all players, or EV, except $i$ ;
\item \textit{Utility}: cost depending on the voltage level
\begin{equation}
\label{PersonalCost}
u_{i}(P_{i},P_{-i})=f^{global,local}(\Delta P) \textrm{.}
\end{equation}
\end{itemize}

We now present the class of potential games, initially introduced by \cite{Monderer1996}, and particularly used in game theory for modeling congestion effects, as in transportation networks for example \cite{Wardrop1952}. A game is said to have a potential, or to be a potential game, if there exists a function $\Phi$ such that 

\begin{eqnarray}
\label{GameOrdPotential}
\forall i \in \mathcal{I}, \, \forall P=(P_{i},P_{-i}), \, \forall P_{i}^{\prime}, & \\
u_{i}(P_{i}^{\prime},P_{-i}) \geq u_{i}(P_{i},P_{-i}) \Leftrightarrow & \Phi(P_{i}^{\prime},P_{-i}) \geq \Phi(P_{i},P_{-i}) \textrm{.} \nonumber
\end{eqnarray}
As in physics, this notion of potential corresponds to a field obtained from an initial problem and from which important properties may be derived. In such a game, a deviation of any player, which makes its utility increase, leads to an increase in the global function $\Phi$. This implies in particular that the asynchronous BRD, used here in the context of voltage regulation, converges because it can not make $\Phi$ infinitly increase during the dynamics.

In this framework, with global objectives, the auxiliary game admits thus $\Phi=f^{global}$ as a potential function and the proposed asynchronous global algorithm will thus converge. This is of practical interest to know that, after a finite time, the updating process will stop and a stable state will be reached. However, neither for the synchronous algorithm, nor for the case with local objectives, this convergence is theoretically proven. But both these approaches must be useful in the practice because a synchronous update of the strategies may permit to accelerate the convergence while the local approach decreases the amount of data to exchange. This will be analyzed by simulations.

\section{Simulation results}

The decentralized methods described in the previous section will be performed using a residential electrical distribution feeder with a large penetration of EV (up to one EV at each node) connected to the standard test network model IEEE $34$ with impedances and lengths scaled down to correspond to a Low-Voltage network and adjusting the R/X ratios, greater in distribution systems. Unless otherwise specified, the number of considered EV will be $30$. Their charging needs are normally distributed with a mean distance of $30$km and a standard deviation of $3$km, taking into account that a $24$kWh battery has a range of $150$km. Concerning the local approach, two neighborhoods are defined: if an EV is connected at node $1$ to $14$ (respectively $15$ to $34$), it only takes into account the voltage on nodes $1-14$ (respectively $15-34$) when deciding its charging policy in the local mechanism. 

\begin{figure}[!htbp]
\centering
\includegraphics[scale=0.32]{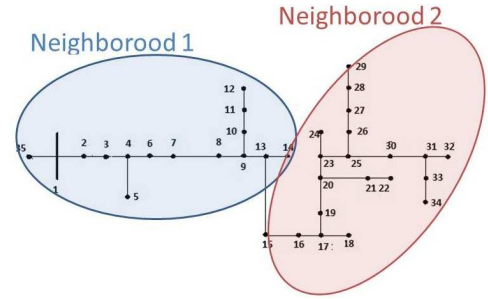}
\caption{IEEE $34$ distribution network model used in the simulations.}
\label{DistribNet}
\end{figure}

\subsection{Convergence of the proposed mechanisms}
 
\begin{figure}[tbp]\vspace{-3mm}
\begin{center}\vspace{-2mm}
\subfigure[Asynchronous approach]{
\resizebox{0.45\columnwidth}{0.46\columnwidth}{\hspace*{-0.7in}\includegraphics{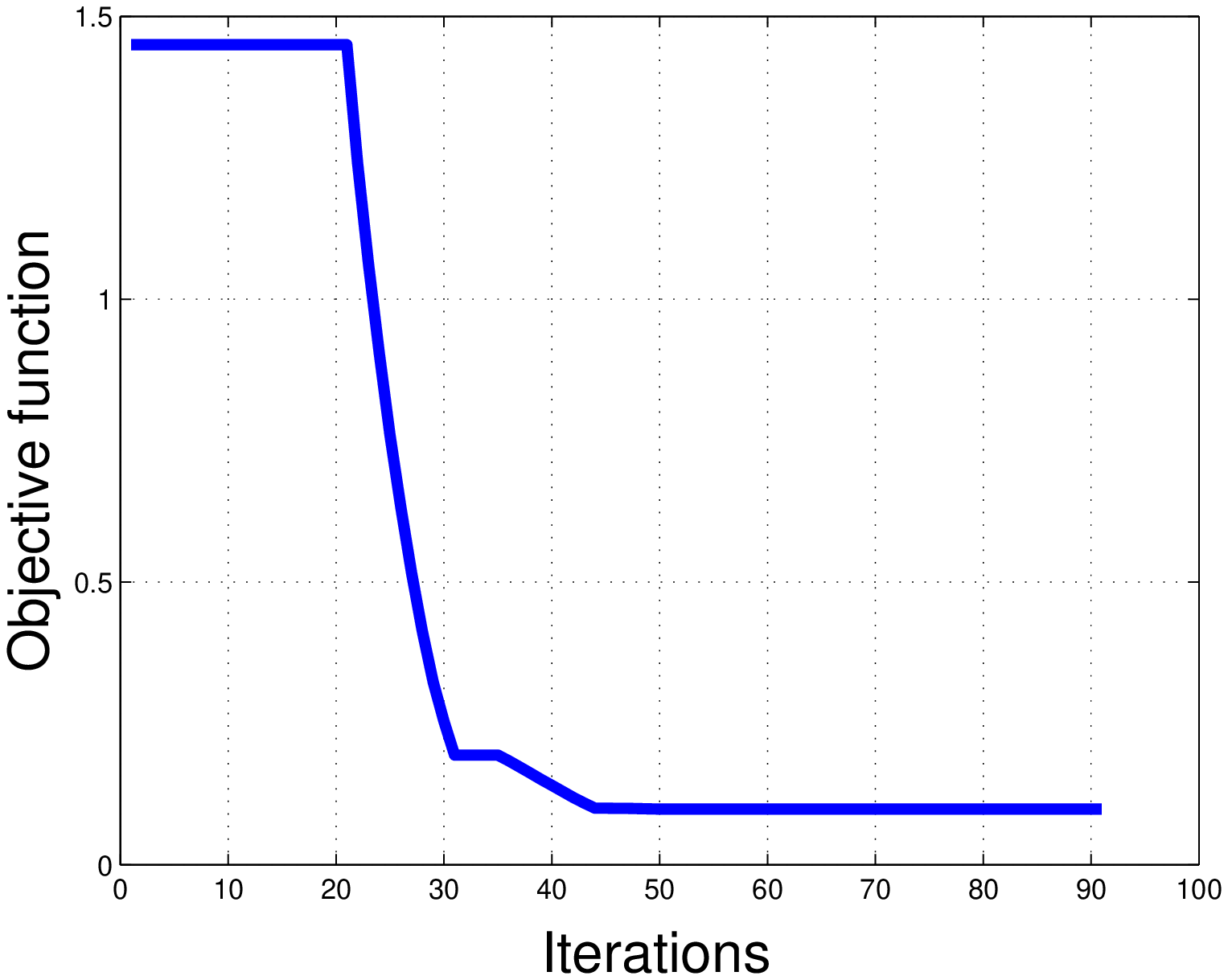}\hspace*{-0.7in}}
\label{fig:Glob}}\vspace{-2mm}
\subfigure[Synchronous approach]{
\resizebox{0.45\columnwidth}{0.46\columnwidth}{\hspace*{-.2in}\includegraphics{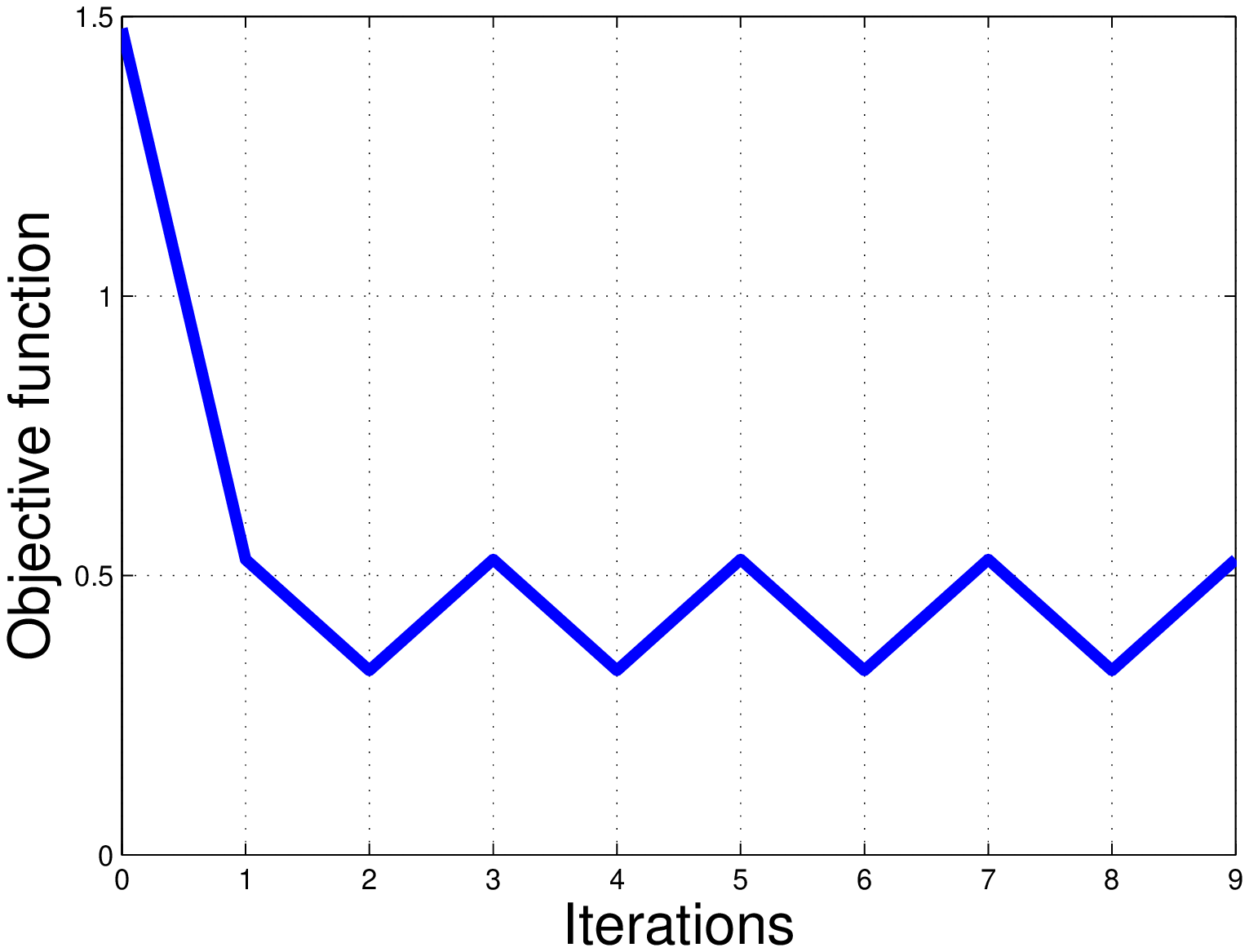}\hspace*{-1in}}\vspace{-2mm}
\label{fig:Loc}} 
\caption{Dynamics of the asynchronous and synchronous global mechanisms}
\label{fig:Dyn} 
\end{center}\vspace{-5mm}
\end{figure}

First, the convergence of the proposed mechanism is analyzed. Given that it is repeated for each time slot, only one particular time slot, $00-00:30$ am, is here presented. As claimed in the previous theoretical part, Fig.\ref{fig:Dyn} shows that, while the asynchronous global algorithm converges, the synchronous global algorithm can have oscillations. Interestingly, convergence in the asynchronous case occurs after approximately $45$ iterations, that is a mean of $1.5$ iterations by EV. This small number of update of EV charging policies is very interesting for practical applications. Note also that, even if the synchronous process does not converge, it provides a very efficient configuration after only one iteration, that is one update by EV. In practice, this scheme could thus be applied using the simple following stopping heuristic rule : when a charging configuration already observed in the coordination mechanism is obtained, the process is stopped. For example, in the case of Fig.\ref{fig:Dyn}, the synchronous mechanism would have been stopped after $3$ iterations. Otherwise, the same cycle will be infinitely repeated without improving the considered objective. Consequently, balancing between obtaining a good state in term of voltage and exchanging little information, the aggregator could thus choose between both approaches. Similar results are observed, but not presented here, in the local case.

%

\subsection{Performance of the proposed methodology}

\begin{figure}[tbp]\vspace{-3mm}
\begin{center}\vspace{-2mm}
\subfigure[Global approach]{
\resizebox{0.45\columnwidth}{0.46\columnwidth}{\hspace*{-0.2in}\includegraphics{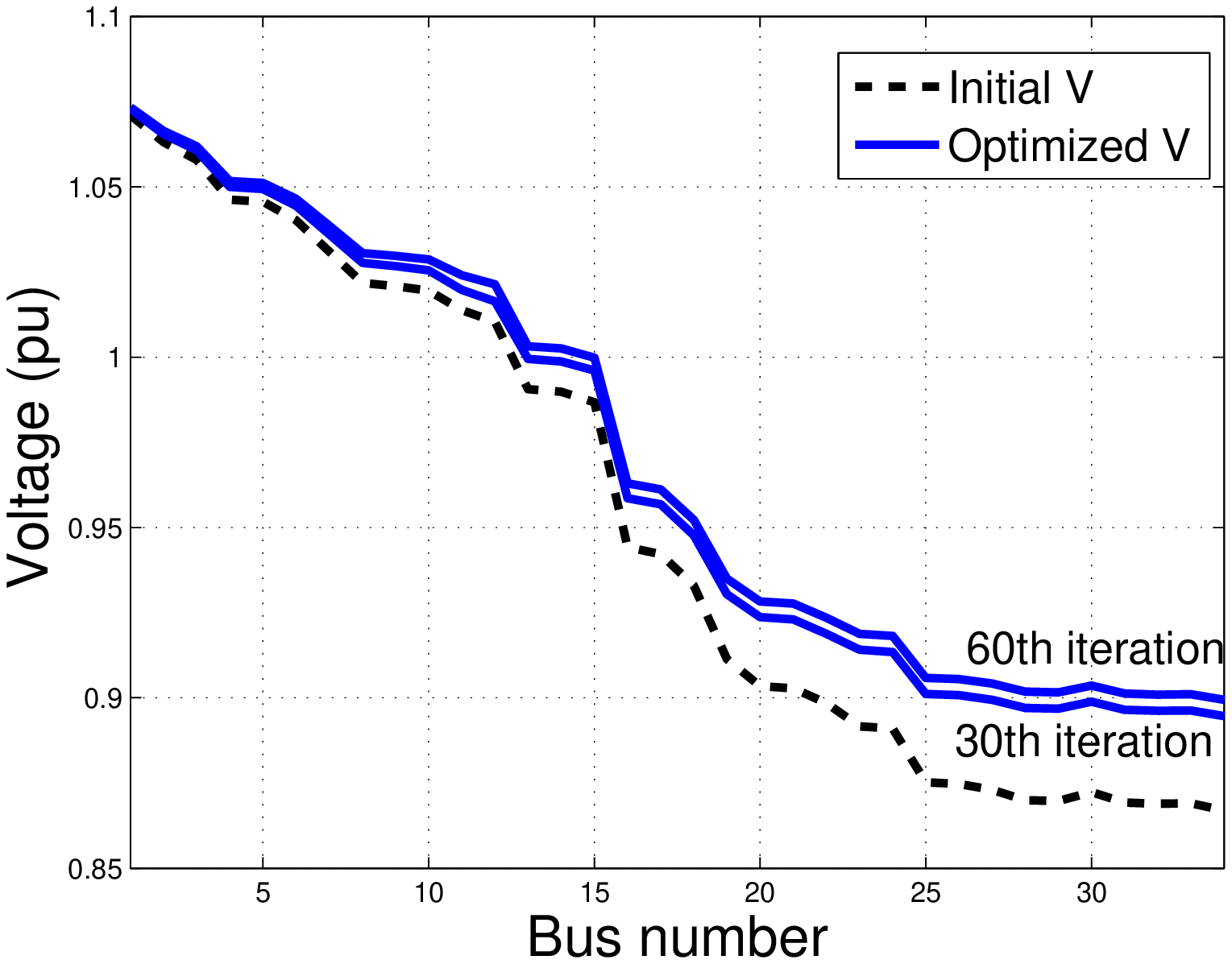}\hspace*{-0.2in}}
\label{fig:Glob}}\vspace{-2mm}
\subfigure[Local approach]{
\resizebox{0.45\columnwidth}{0.46\columnwidth}{\hspace*{-.2in}\includegraphics{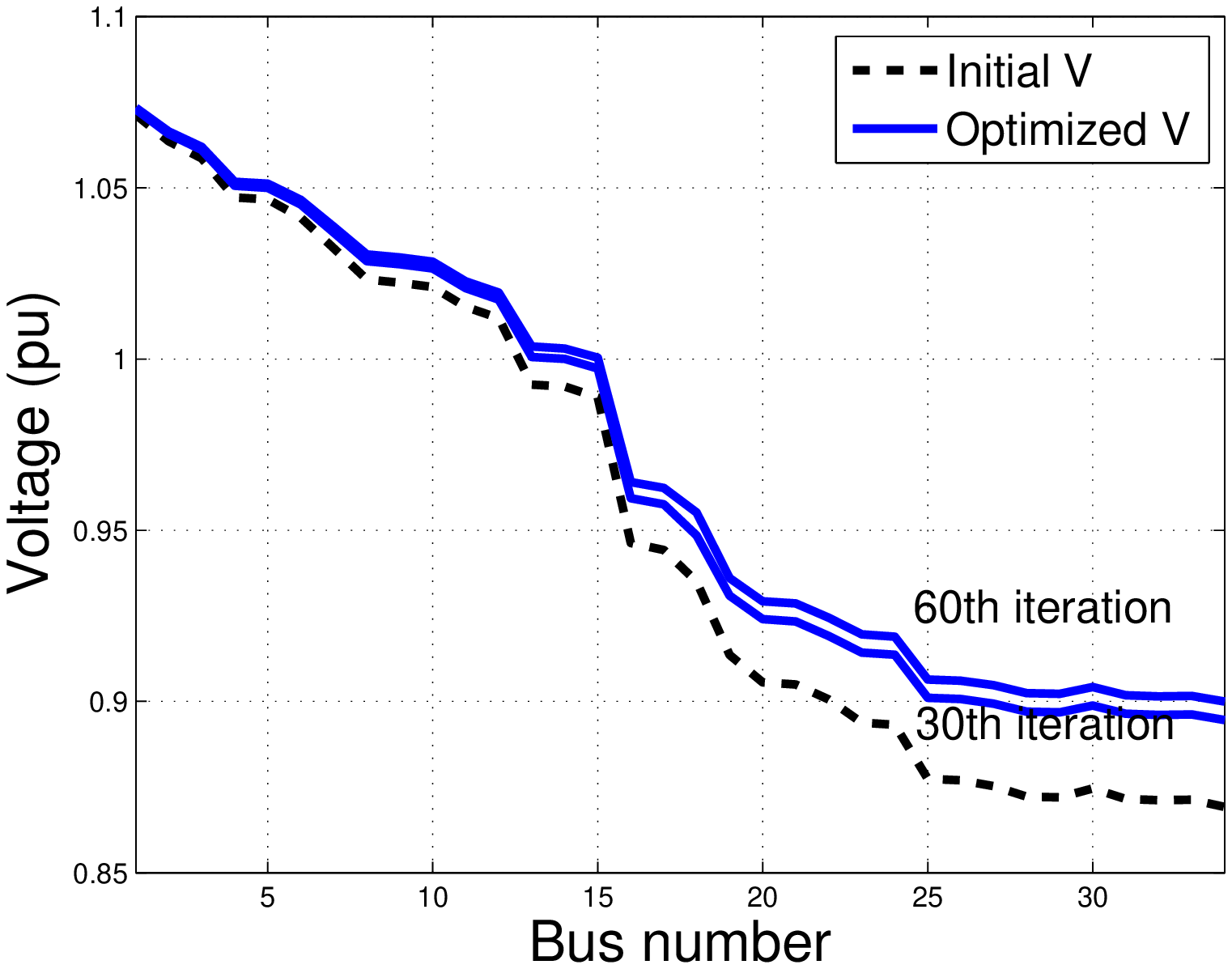}\hspace*{-1in}}\vspace{-2mm}
\label{fig:Loc}} 
\caption{Voltage profile obtained with the global and local mechanisms}
\label{fig:VProf} 
\end{center}\vspace{-5mm}
\end{figure}

Having ensured that after a finite (short) time, this mechanism leads to a stable configuration, then, its performance is studied observing the voltage profile, especially at the node $34$ at the end of the considered network topology, probably the one which will be the more significantly impacted by an uncoordinated EV charging. Fig.\ref{fig:VProf} shows that, without coordinating EV charging, the voltage at the end of the network is under the limit of $0.9$ pu. With both the global and the local approaches, the voltage remains in the standard limits. Both profiles are very similar with only a small difference for the voltage of the nodes $7-14$, which are taken into account differently in both methods. Interestingly, the local mechanism seems here sufficient to coordinate efficiently EV charging. 

%

\medskip

Then, the performance of the proposed methodology is compared with two other scenarios :
\begin{itemize}
\item without EV charging regulation, called \textit{uncoordinated case}. EV charge as soon as connected to the grid, with normal distributions for arrival time of mean $6:45$ pm and standard deviation $1$ hour and for departure time of mean $8$ am and standard deviation $0.75$ hour. This uncoordinated case is the one where the voltage may be the more adversely affected ;
\item with a \textit{voltage droop charging control}, as proposed in \cite{Geth2012}. Given the voltage at time $t$ at the node where EV is located, the EV charges at time $t+1$ according to the profile of Fig.\ref{fig:droopContr} (three are given and compared in \cite{Geth2012}, but ony the so-called "LM1" is considered here). 
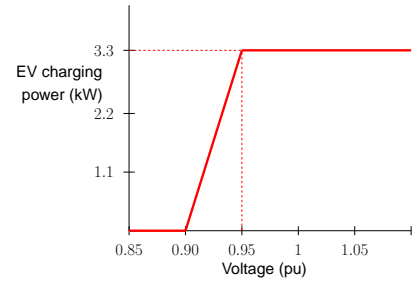
\begin{figure}[!htbp]
\begin{center}
\scalebox{0.3}{
\begin{tikzpicture}[y=10cm,x=.5cm,font=\sffamily]
 	
	\draw[line width=1] (0,0.2) -- (25,0.2);
	\draw[line width=1] (0,0.2) -- (0,1.2);
	\draw[line width=1] (0,0.18) -- (0,0.22);	
	\draw[line width=1] (5,0.18) -- (5,0.22);	
	\draw[line width=1] (10,0.18) -- (10,0.22);	
	\draw[line width=1] (15,0.18) -- (15,0.22);	
	\draw[line width=1] (20,0.18) -- (20,0.22);	
	\draw[line width=1] (25,0.18) -- (25,0.22);	
	\draw[line width=1] (-0.5,1) -- (0.5,1);
	\draw[line width=1] (-0.5,0.46) -- (0.5,0.46);
	\draw[line width=1] (-0.5,0.72) -- (0.5,0.72);
	
	\node[below=0.3cm] at (0,0.18) {\huge $0.85$};
	\node[below=0.3cm] at (5,0.18) {\huge $0.90$};
	\node[below=0.3cm] at (10,0.18) {\huge $0.95$};
	\node[below=0.3cm] at (15,0.18) {\huge $1$};
	\node[below=0.3cm] at (20,0.18) {\huge $1.05$};
	\node[left=0.3cm] at (-0.5,0.46) {\huge $1.1$};
	\node[left=0.3cm] at (-0.5,0.72) {\huge $2.2$};
	\node[left=0.3cm] at (-0.5,1) {\huge $3.3$};
	\node[below=0.3cm] at (12,0.1) {\huge Voltage (pu)};
	\node[left=0.3cm] at (-1.5,0.9) {\huge EV charging};
	\node[left=0.3cm] at (-1.5,0.8) {\huge power (kW)};
	\draw[color=red,line width=3] (0,0.2) -- (5,0.2);
	\draw[color=red,line width=3] (5,0.2) -- (10,1);
	\draw[color=red,line width=3] (10,1) -- (25,1);
	\draw[red,dashed] (10,0.2)--(10,1);
	\draw[red,dashed] (0,1)--(10,1);
\end{tikzpicture}
}
\caption{Voltage droop charging control of EV charging proposed in \cite{Geth2012}.
}
\label{fig:droopContr}
\end{center}
\end{figure}
\end{itemize}  

\begin{figure}[!htbp]
\includegraphics[scale=0.45]{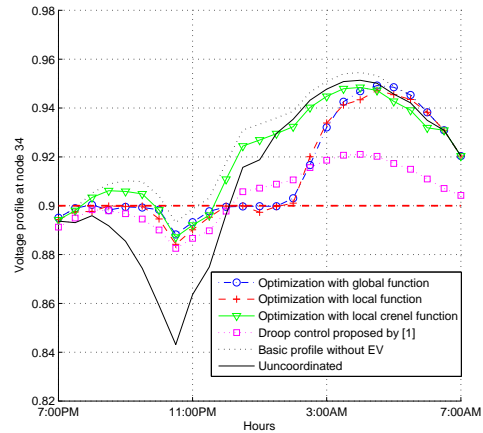}
\caption{Voltage profile obtained with different charging mechanisms.}
\label{Fig:VDiffApp}
\end{figure}

Fig.\ref{Fig:VDiffApp} shows that an uncoordinated EV charging would lead to a critical voltage drop at the end of the evening. With the proposed decentralized mechanisms, this is significantly reduced and the voltage remains almost within its standard bounds, which is also the case with the droop charging control. Moreover, observe that when the voltage is between $0.9$ and $1.1$ pu, the performance of the decentralized control with the quadratic or crenel objective and the droop voltage control can not be distinguished given that the metrics $f_V$ used here to quantify voltage deviations is zero within this interval. Furthermore, the results with both the quadratic or crenel metrics seems very similar. An extension of this initial work could be to analyze in more details the sensibility of the results obtained with respect to the choice of the metrics used.

Finally, the different mechanisms are compared over a wide range of configurations, differentiated by the localizations of the EV which are randomly drawn ($10$ random drawings of the localization are done). To this end, Tab.\ref{tab:perfV} provides the mean of the minimal voltage over the day at node $34$. With this criterion, the proposed decentralized methodology slightly outperforms the voltage droop charging method and significantly the uncoordinated case. The decentralized local mechanism seems also to provide similar results than the global approach. Note finally that the more EV connected to the network is, the bigger the impact on the voltage of node $34$ is, which is very intuitive. 

\begin{figure}[!htbp]
\centering
\includegraphics[scale=0.3]{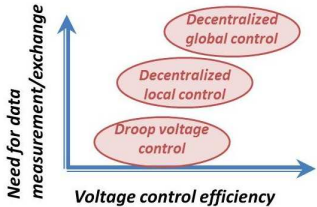}
\caption{First overview of the characteristics of the different EV charging control methods}
\label{Fig:PerfDiffMeth}
\end{figure}

To summarize the efficiency of the different mechanisms analyzed in this paper, Fig.\ref{Fig:PerfDiffMeth} outlines their characteristics according to two key aspects : efficiency in terms of voltage control and need for data exchange for these methods to be applied. With the simulations done here, there is only a slight difference according to the voltage control and choice could be made to apply the mechanism with the smallest need for data exchange. Further simulations on other type of networks could put into question this first conclusion.


\begin{table}[h]
\caption{\textbf{Comparison of the minimal voltage at node $34$ with different EV charging control mechanisms}}
\label{tab:perfV}
\begin{center}
\begin{tabular}{|c|c|c|c|}
\hline
Policy \& Number of EV & 10 & 20 & 30 \\
 \hline
Uncoordinated &	0.877 & 0.866 & 0.849 \\
\hline
Droop voltage control &	0.886 & 0.881 & 0.883 \\
\hline
Global asynchronous decentralized control &	0.892 & 0.890 & 0.888 \\ 
\hline
Local asynchronous decentralized control & 0.891 & 0.888 & 0.885 \\
\hline
\end{tabular}
\end{center}
\end{table}

\section{Concluding remarks}

In this paper, we proposed a decentralized approach for scheduling EV charging in a distribution network in order to reduce its impact on the voltage plan. Concretely, this consists in an iterative exchange of information between a coordinator, or aggregator, and EV connected in the district until a stable configuration is reached. Receiving information from the aggregator, each EV aims at limiting its impact on the voltage plan using the sensitivity matrix approach in order to estimate the voltage on the key points of the network, or pilot nodes, according to its charging policy. More precisely, two different iterative algorithms are defined, namely a global and local mechanisms. In the global method, each EV takes into account its impact on the whole network when updating its charging choice while in the local one it only focuses on its impact on a close neighborhood of its connection node, assuming that its influence will be the more significant in this subpart of the network. Asynchronous and synchronous update of EV charging decisions were also studied. 

Using properties from game theory, the global approach is shown to converge when EV update their decisions asynchronously, which is of practical interest for a concrete application. Then, realistic simulations on an IEEE test network show the significant benefits done when applying these two mechanisms in comparison with the uncoordinated case and in comparison with a voltage local droop control recently proposed in the literature. This analysis also highlights the respective advantages of asynchronous and synchronous mechanisms: while the first permits to obtain a better configuration in terms of voltage plan, the second can reach an efficient configuration with significantly less iterations, which means less information exchanged between the aggregator and the EV. The local approach is also shown to provide very similar results in comparison with the global approach, for which more data has to be exchanged. Between both of these methods, a mixed approach, where subdistricts exchange information via aggregators and then a local decentralized mechanism is applied in each subdistrict, constitutes a relevant extension of this work.
\ifCLASSOPTIONcaptionsoff
  \newpage
\fi

\bibliographystyle{IEEEtran_NoURL}
\bibliography{LittISGT_2013_V3}

\end{document}